\documentclass[useAMS,usenatbib]{mn2e}

\usepackage{epsf}

\usepackage{color}
\definecolor{Blue}{cmyk}{1.0,1.0,0.0,0.0}
\definecolor{Green}{cmyk}{1.0,0.0,1.0,0.0}
\definecolor{Orange}{cmyk}{0.0,0.5,1.0,0.0}
\definecolor{Red}{cmyk}{0.0,1.0,1.0,0.0}

\def\lsim{\lower 2pt \hbox{$\, \buildrel {\scriptstyle <}\over
{\scriptstyle \sim}\,$}}
\def\gsim{\lower 2pt \hbox{$\, \buildrel {\scriptstyle >}\over
{\scriptstyle \sim}\,$}}

\title[3-D Radiative Properties of Hot Accretion
  Flows]{Three-dimensional Radiative Properties of Hot Accretion Flows
  onto the Galactic Centre Black Hole}
\author[Y. Kato, M. Umemura, K. Ohsuga]{Y. Kato$^{1}$\thanks{E-mail:
    kato.yoshiaki@isas.jaxa.jp}, M. Umemura$^{2}$, K. Ohsuga$^{3}$\\
$^{1}$Institute of Space and Astronautical Science, Japan Aerospace
Exploration Agency, 3-1-1 Yoshinodai, Sagamihara, \\
Kanagawa 229-8510, Japan\\
$^{2}$Center for Computational Sciences, University of Tsukuba, 
1-1-1 Ten-nodai, Tsukuba 305-8577, Japan\\
$^{3}$National Astronomical Observatory of Japan, Osawa, Mitaka, Tokyo
181-8588, Japan}
\begin{document}

\date{Accepted 1988 December 15. Received 1988 December 14; in original form 1988 October 11}

\pagerange{\pageref{firstpage}--\pageref{lastpage}} \pubyear{2002}

\maketitle

\label{firstpage}

\begin{abstract}
By solving radiative transfer equations, we examine three-dimensional
radiative properties of a magnetohydrodynamic accretion flow model
confronting with the observed spectrum of Sgr A*, in the vicinity of
supermassive black hole at the Galactic centre. As a result, we find
that the core of radio emission is larger than the size of the event
horizon shadow and its peak location is shifted from the gravitational
centre.  We also find that the self-absorbed synchrotron emissions by
the superposition of thermal electrons within a few tens of the
Schwartzschild radius can account for low-frequency spectra below the
critical frequency $\nu_{c}\approx 10^{12}$ Hz.  Above the critical
frequency, the synchrotron self-Compton emission by thermal electrons
can account for variable emissions in recent near-infrared
observations.  In contrast to the previous study by Ohsuga et
al. (2005), we found that the X-ray spectra by Bremsstrahlung emission
of thermal electrons for the different mass accretion rates can be
consistent with both the flaring state and the quiescent state of Sgr
A* observed by {\it Chandra}.
\end{abstract}


\section{Introduction}
How the emission comes from accreting material in the Galactic centre
(GC) is a fundamental question for understanding the nature of mass
accretion processes feeding a supermassive black hole (SMBH).  When
mass accretion rate is much less than the critical value,
$\dot{M}_{\rm crit}\equiv L_{\rm Edd}/c^{2}$ where $L_{\rm Edd}\approx 1.3\times 10^{38} \left(M_{\rm BH}/M_{\odot}\right){\rm
  erg\,s^{-1}}$ is the Eddington luminosity and $c$ is the speed of
the light, the radiation loss of accreting gas is inefficient and thus
most of the energy generated by turbulent viscosity is stored as
thermal energy of the gas and is advected onto SMBH.  Therefore, the
accretion flow becomes hot and geometrically thick structure. This
type of accretion flow is well known as an advection-dominated
accretion flow (ADAF: Ichimaru 1977; Narayan~\& Yi 1994, 1995;
Abramowicz et al. 1995) or a radiatively inefficient accretion flows
(RIAF: Yuan et al 2003; see also Kato, Fukue, Mineshige 2008).  The
ADAF/RIAF model is quite successful in reproducing high-energy
emission of low-luminous active galactic nuclei (AGN) and the GC
source, which is Sgr A* as a compact radio source (Balick~\& Brown
1974; Bower et al. 2004; Shen et al. 2005; Doeleman et al. 2008).
Actually, the stellar dynamics has revealed the mass of SMBH in the GC,
$\approx 4\times 10^{6} M_{\odot}$ (e.g., Sch{\"o}del et al. 2002; Ghez
et al. 2003, 2008; Gillessen et al. 2009).  It turned out that the
low-luminous material at the GC is associated with Sgr A*, in which
the luminosity is $L_{\rm Sgr A*}\approx 10^{-8} L_{\rm Edd}$.

After the discovery of magneto-rotational instability (MRI: Balbus~\&
Hawley 1991, 1998), which can drive magnetohydrodynamic (MHD) turbulence as
a source of viscosity for accretion process, it seems that
non-radiative MHD simulations of accretion flows have been accepted as
a realistic model of ADAF/RIAF (Stone~\& Pringle 2001; Hawley~\&
Krolik 2001).  For example, many MHD studies based on numerical
simulations have revealed a hot and geometrically thick accretion
flows with a variety of complex motions (Matsumoto et al. 1999; Hawley
2000; Machida et al. 2000), outflows and jets (Igumenshchev et
al. 2003; Proga~\& Begelman 2003; Kato et al. 2004), and oscillations
(Kato 2004).  The recent detection of variability in the X-ray and
near-infrared (NIR) emissions at Sgr A* (Baganoff et al. 2001, 2003;
Ghez et al. 2004) may be induced by such a multi-dimensional structure
of the flow.  Nonlocal nature of radiation process is essential for
testing MHD models.  Therefore, in order to clarify the structure and
the time variability of accretion flows, undoubtedly full radiation
transfer treatment of MHD accretion flows in three-dimension is
indispensable.

The pioneering work for examining MHD model of accretion flows has
been done by Ohsuga, Kato, Mineshige (2005: hereafter OKM05).  OKM05
assumed the cylindrical distribution of electron temperature by
adopting the balance equation between radiative cooling and heating
via Coulomb collision, regardless of gas temperature in MHD model.
They reconstructed for the first time the multi-band spectrum of MHD
model which is consistent with the observed spectra in flaring state
of Sgr A*.  However, the spectra in quiescent state cannot be
reconstructed simultaneously in the radio and X-ray bands.  In the
context of MHD model, this is the issue of what makes the difference
between the flaring state and the quiescent state in Sgr A*.  We expect
that determination of the electron temperature is the key for
understanding the occurrence of two distinct states.

Moscibrodzka, Proga, Czerny, and Siemiginowska (2007: hereafter
MPCS07) present the spectral feature of axisymmetric MHD flows by
Proga~\& Begelman (2003).  In contrast to OKM05, they calculate the
electron temperature distribution by solving the heating-cooling
balance equation at each grid point at a given time in the simulation.
Moreover, they take into account the advective energy transport and
compressive heating in the balance equation.  It turns out that the
heating of electrons via Coulomb collision is not always a dominant
term in the balance equation.  Unfortunately, MPCS07 failed to
reproduce the radio and X-ray spectra by emission from thermal
electron.  They conclude that a contribution of non-thermal electrons
offers a much better representation of the spectral variability of Sgr
A*.

One difficulty in theoretical studies of radiative feature of
hot accretion flows is that radiative energy transfer plays a critical
role for determining the electron temperature.  For example, radiative
heating/cooling via synchrotron emission/absorption and Compton
processes may dominate compressional heating and collisional heating
in the energy balance equation at the high electron temperature (Rees
et al. 1982).  This makes everything rather complicated than the case
of MPCS07.  Conversely, once we know the properties of two-temperature
plasma in MHD accretion flows which can reconstruct a radiation
spectrum using a few relevant parameters, we can then constrain the
physics of heating mechanism.

In this study, we investigate radiative signatures of radiatively
inefficient accretion flows in long-term 3-D global MHD simulations.
Then, confronting with the observed spectra of Sgr A* in the flaring
and quiescent states, we derive constraints for electron temperature
that can concordant with the observations.  We also discuss heating
mechanism of electron in the magnetized accretion flows.  In \S 2 we
present our 3-D MHD model of accretion flows and describe method of
radiation transfer calculation. We then present our results in \S
3. The final section is devoted to summary and discussion.

\section{Numerical Models and Methods}

\subsection{Setup of simulations}
A physical model we use in this study is based on 3-D resistive MHD
calculations with pseudo-Newtonian potential (see Kato, Mineshige,
Shibata 2004; Kato 2004 for more details).  We pick last 40
snapshots of our calculations from $t=30,000\,r_{\rm s}/c$ to
$32,000\,r_{\rm s}/c$ with $dt=50 r_{\rm s}/c$ where $r_{\rm s}$ and
$c$ are the Schwartzschild radius and the speed of light,
respectively.  Note the flow has evolved in more than 600 rotation
periods at the innermost stable circular orbits (ISCO) where the
Keplerian rotation period is about $50 r_{\rm s}/c$.  Therefore our
MHD model is supposed to be in quasi-steady state.  In our radiation
transfer calculations, we use uniform meshes as $(N_{\rm x},N_{\rm
  y},N_{\rm z}) = (100, 100, 100)$ in Cartesian coordinates in a
simulation box $|x, y, z| \leq 100\,r_{\rm s}$.

\subsection{MHD model}
\begin{figure*}
  \centerline{\epsfxsize=\hsize \epsfbox{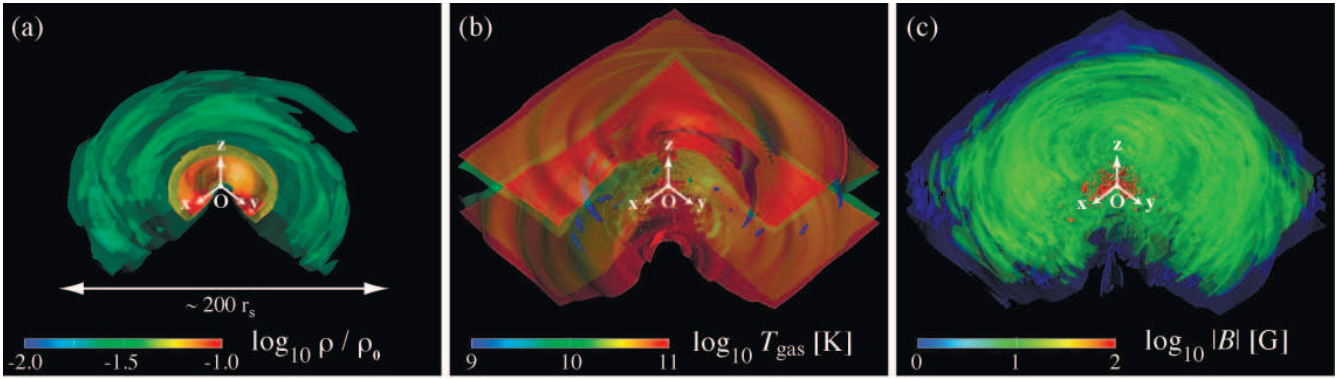}}
  \caption{Perspective view of spatial distributions of density
    (left), gas temperature (middle), and strength of magnetic field
    (right) at $t=30000\,[r_{\rm s}/c]$.  Normalized density is
    $\rho_{0}=8\times 10^{-15}\,{\rm [g\, cm^{-3}]}$.}
  \label{fig1:eps}
\end{figure*}

The quasi-steady MHD accretion discs have a hot, geometrically thick
structure associated with sub-thermal magnetic field.
Fig. \ref{fig1:eps} displays a spatial distribution of the density,
the gas temperature, and the strength of magnetic field of our MHD
model.  The density is normalized by the initial maximum density
$\rho_{0}$ and the strength of magnetic field is proportional to
$\left(\rho/\rho_{0}\right)^{1/2}$ for the same black hole mass
$M_{\rm BH}$.  As we can notice, an MHD disc has non-axisymmetric
structures (close to $m=1$ where $m$ is the azimuthal mode number) in
density, gas temperature, and strength of magnetic field,
simultaneously.  Moreover, filamentary structures of cold and dense
gas can be seen in left and middle panels.  Note that gas temperature
is relatively high at funnel region along the z-axis because the
centrifugal barriers prevent the penetration of accreting material. In
right panel, MHD turbulence induced by MRI and the differential
rotation in the flow generate strong magnetic field regions within
approximately $30 r_{\rm s}$.

  
\subsection{Electron temperature}
In our MHD model, it is assumed that the entire magnetic energy
released in the diffusion region is thermalized instantly and
therefore the production of non-thermal electrons accelerated by the
magnetic reconnection is neglected for self-consistency.  In the
previous study OKM05, electron temperature is determined by the local
thermal equilibrium between radiative cooling and electron heating via
coulomb coupling in the cylindrical region. However, this assumption
is invalid.  Actually, in our preliminary radiation transfer
calculation coupled with the energy balance equation, we found that
the heating rate of coulomb coupling cannot afford the cooling rate of
electrons in each computational cell. Therefore, the other heating
mechanism of electrons, such as turbulent heating, must be taken into
account for physical reasoning. Similar conclusions have been made by
Sharma, Quataert, Hammett,~\& Stone (2007: hereafter SQHS07).  For
this reason, we introduce a new parameter, $f_{\rm ep}\equiv T_{\rm
  e}/T_{\rm  p}$, the ratio of electron temperature and proton
temperature so that electron temperature is determined with the gas
temperature, $T_{\rm gas}$, by using $f_{\rm ep}$ in this study.  We
assume that $f_{\rm ep}$ is spatially uniform for simplicity.  Thus
electron temperature is obtained as:
\begin{equation}
T_{\rm e} = \min{\left(\frac{f_{\rm ep}}{1+f_{\rm ep}}T_{\rm gas},m_{\rm
  e}c^2/k_{\rm b}\right)}
\end{equation}
where $T_{\rm gas}$ is the gas temperature which is derived by MHD
simulation.  We assume here that the electron temperature cannot
exceed the temperature of rest mass energy of electron due to pair
annihilation.

\subsection{Parameters for radiation transfer calculation}
There are only three model parameters, $\rho_{0}$, $f_{\rm ep}$, and
$M_{\rm BH}$, in order to perform radiative transfer calculation in
our study.  The mass of SMBH in the GC is fixed at $M_{\rm
  BH}=3.6\times 10^{6} M_{\odot}$ (e.g., Sch{\"o}del et al. 2002).
The other model parameters we choose here are as follows: (a)
$\rho_{0}=8\times 10^{-15} {\rm g\,cm^{-3}}$ and $f_{\rm ep}=0.25$,
(b) $\rho_{0}=8\times 10^{-15} {\rm g\,cm^{-3}}$ and $f_{\rm ep}=1$,
(c) $\rho_{0}=8\times 10^{-16} {\rm g\,cm^{-3}}$ and $f_{\rm
  ep}=0.25$, and (d) $\rho_{0}=8\times 10^{-16} {\rm g\,cm^{-3}}$ and
$f_{\rm ep}=1$.  Note that these parameters are derived by fitting the
observed broadband spectra.  Because the flow velocity is identical in
all models, the larger density represent the more mass accretion rate
$\dot{M}$.  The relation between $\rho_{0}$ and $\dot{M}$ at ISCO can
be described as follows:
\begin{equation}
\dot{M}\approx 2.5\times 10^{-7}\left({\rho_{0}\over 8\times
  10^{-15}\,{\rm g\,cm^{-3}}}\right) M_{\odot}\,{\rm yr}^{-1}.
\end{equation}
This relation indicates that mass accretion rate of our models is
smaller than that for the quiescent X-ray emission measured with {\it
  Chandra} of $\dot{M}\sim 10^{-6}\,M_{\odot}\,{\rm yr}^{-1}$ at the
Bondi radius (Baganoff et al. 2003), but is consistent with that
estimated by the Faraday rotation in the millimeter band of
$\dot{M}\sim 10^{-7} - 10^{-8}\,M_{\odot}\,{\rm yr}^{-1}$ (Bower et
al. 2003, 2005).

In radiation transfer calculation, we treat synchrotron
emission/absorption, (inverse-)Compton scattering, and bremsstrahlung
emission/absorption of the thermal electrons.  Non-thermal electrons
produced by collisions of protons via $\pi$-decay (Mahadevan et
al. 1998) are not taken into account, because we focus only on the
radiative properties of thermal electrons (Loeb~\& Waxman 2007).

\subsection{Radiative transfer calculation}
\begin{figure}
  \centerline{\epsfxsize=0.9\hsize \epsfbox{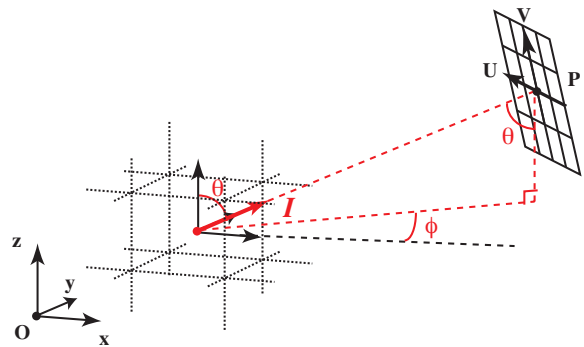}}
  \caption{Geometry of radiation fields,
    $I$ in the coordinate space $(x,y,z,\theta,\phi)$ and coordinate
    screen $(u,v)$ for the ray-tracing calculation.}
  \label{fig2:eps}
\end{figure}

We solve the following radiation transfer equations with electron
scattering by using Monte-Carlo method:
\begin{equation}
\mbox{\boldmath$n$}\cdot\nabla\mbox{\boldmath$I$}_{\nu} =
\chi_{\nu}(\mbox{\boldmath$S$}_{\nu} - \mbox{\boldmath$I$}_{\nu})
\end{equation}
where
$\mbox{\boldmath$I$}_{\nu}(x,y,z,\theta,\phi)$ is the specific
intensity at the position $(x,y,z)$ in the direction $(\theta,
\phi)$ with the frequency $\nu$, whereas
$\chi_{\nu}(x,y,z,\theta,\phi) = n_{\rm e}\sigma_{\nu} + \kappa_{\nu}$
is the extinction coefficient where $n_{\rm e}$, $\sigma_{\nu}$ and
$\kappa_{\nu}$ is the electron number density, the scattering
cross-section, and the absorption coefficient, respectively, and
\begin{equation}
\mbox{\boldmath$S$}_{\nu} = {\varepsilon_{\nu}\over 4\pi\chi_{\nu}} +
\oint\varphi(\nu, \mbox{\boldmath$n$}; \nu', \mbox{\boldmath$n$}')
\alpha_{\nu'}\mbox{\boldmath$I$}_{\nu'}(\mbox{\boldmath$n$}')d\mbox{\boldmath$\Omega$}'
\end{equation}
is the source function where $\varepsilon_{\nu}$,
$\alpha_{\nu}$, and $\varphi(\nu, \mbox{\boldmath$n$}; \nu'
\mbox{\boldmath$n$}')$ is the local emissivity, the scattering albedo,
and the phase function, respectively.  The coordinate system is shown
in Fig. \ref{fig2:eps}.

For generating photon packets, we randomly selected a position by
using a local emissivity as follows:
\begin{equation}
\sum_{\rm i=1}^{\rm k-1}\varepsilon_{\nu}({\rm i}) < R_{1}\sum_{\rm
  i=1}^{\rm Nmesh}\varepsilon_{\nu}({\rm i}) < \sum_{\rm i=1}^{\rm
  k}\varepsilon_{\nu}({\rm i})
\end{equation}
where $\varepsilon_{\nu}({\rm i})$ is the emissivity at the position
index ${\rm i}$ and $R_{\rm 1}$ indicates a random number distributed
uniformly in the interval $[0, 1]$.  In our study, all pseudo-random
numbers $R_{\rm j}$ are generated by using a Mersenne Twister method
(Matsumoto~\& Nishimura 1998).  The direction of generated
photon packets, $\mbox{\boldmath$n$}=(\sin{\theta}\cos{\theta},
\sin{\theta}\sin{\phi},\cos{\theta})$, is also determined by using
random numbers $R_{2}=(\cos{\theta} + 1)/2$, $R_{3}=\phi/2\pi$.
The frequency domain of photon packets is ranging from $10^{3}$ to
$10^{25}$ Hz and is uniformally divided by 100 bins in logarithmic
scale.  In this study, $N_{\rm p} = 10^{7}$ photon packets are
generated in every frequency bin in order to acquire the statistically
significant results.

In order to evaluate an escaping probability of an emerging photon
packet, optical depth is computed by using a direct integration along
the photon packet trajectory as follows:
\begin{equation}
\tau_{\nu}(l)=\int_{0}^{l} \left(n_{\rm e}\sigma_{\nu} + \kappa_{\nu}\right)ds,
\end{equation}
where $l$ is a distance between the origin of the photon packet and
the computational boundary along the trajectory.  Here, we use the
Klein-Nishina formula of scattering cross-section $\sigma_{\rm KN}$ for
$\sigma_{\nu}$ (Rybicki~\& Lightman 1979) and the synchrotron,
free-free, and bound-free self-absorption coefficient for
$\kappa_{\nu}$  described by Kirchoff's law assuming local thermal
equilibrium (LTE) at every meshes,
\begin{equation}
\kappa_{\nu}=\frac{\varepsilon^{\rm sy}_{\nu} + \varepsilon^{\rm
    ff}_{\nu} + \varepsilon^{\rm bf}_{\nu}}{4\pi B_{\nu}}
\end{equation}
where $\varepsilon^{\rm sy}_{\nu}$, $\varepsilon^{\rm ff}_{\nu}$, and
$\varepsilon^{\rm bf}_{\nu}$ are synchrotron, free-free, and
bound-free emissivity, respectively (Pacholczyk 1970; Stepney~\&
Guilbert 1983), and $B_{\nu}$ is the Planck function.  Accordingly, an
escaping probability of the generated photon packet is written as:
\begin{equation}
w(l)=\exp{(-\tau_{\nu}(l))},
\end{equation}
and the remaining photon packet $1 - w(l)$ interacts with
gas. Scattering position at the distance $l_{\rm s}$ from the original
position is determined by using a random number as follows:
\begin{equation}
R_{4} = \left[1 - w(l_{\rm s})\right]/\left[1 - w(l)\right]
\end{equation}
and the scattering albedo is given by:
\begin{equation}
\alpha_{\nu} = \frac{n_{\rm e}\sigma_{\nu}}{n_{\rm e}\sigma_{\nu} +
  \langle\kappa_{\nu}\rangle}
\end{equation}
where $\langle\kappa_{\nu}\rangle$ is the mean absorption
coefficient along the photon packet trajectory.  To determine the
phase function for the Compton scattering process,
$\varphi(\nu, \mbox{\boldmath$n$}; \nu', \mbox{\boldmath$n$}')$, we
follow the method by Pozdnyakov et al. (1977).  We repeat the same
procedure for a scattered photon packet until either a photon is out
of the computational box or $w(l) < \epsilon$ where $\epsilon=10^{-5}$
in this study.  Note that photons can neither penetrate nor be
generated at the region within 2 $r_{\rm s}$ in above calculation.

Finally, the computed radiation field is accumulated in the data array
of $(N_{\rm x},N_{\rm y},N_{\rm z},N_{\theta},N_{\phi},N_{\nu}) =
(100, 100, 100, 3, 4, 5)$ in order to generate synthetic images in a
given frequency band and viewing angle.  On the other hand, the
computed spectrum of escaping photons in the fixed viewing angle is
stored in the different data array.

\section{Results}

\subsection{Mapping of escaping photons}
\begin{figure*}
  \centerline{\epsfxsize=0.9\hsize \epsfbox{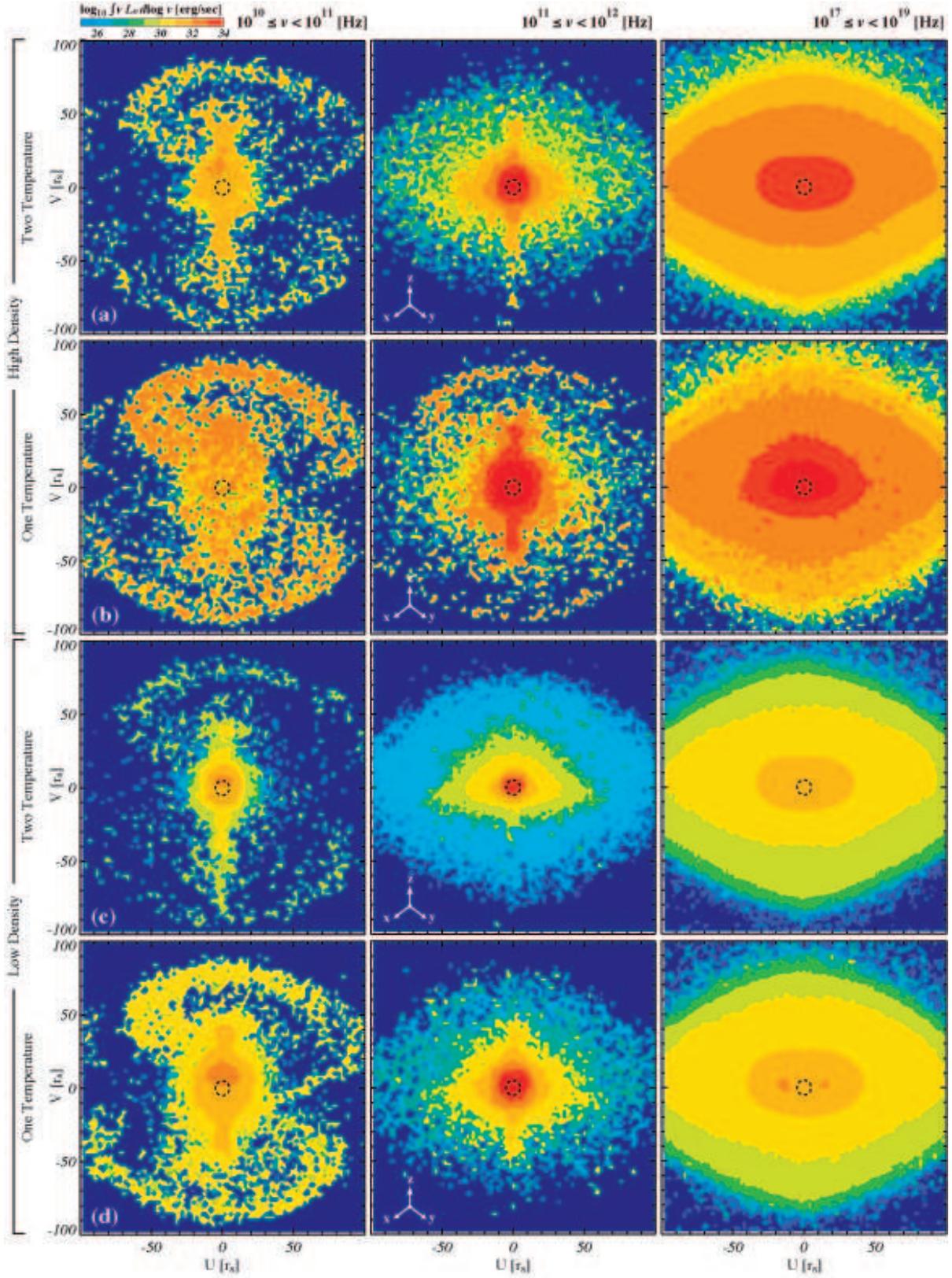}}
  \caption{Synthetic images of four models [(a), (b), (c), and (d)
      from top to bottom] in different frequencies (left: $10^{10}\leq
    \nu < 10^{11}$ [Hz]; centre: $10^{11}\leq\nu < 10^{12}$ [Hz];
    right: $10^{17}\leq\nu <10^{19}$ [Hz]) at the viewing angle of
    ($\theta$, $\phi$)$=$($\pi/3$, $\pi/4$).  Colours indicate the
    power of escaping photon packets.  A dotted circle with the radius
    $5 r_{\rm s}$ indicates the canonical size of event horizon
    shadow.}
  \label{fig3:eps}
\end{figure*}

We investigate the spatial distribution of emergent radiation in
order to explore the radiative nature of magnetized accretion flows.
Fig. \ref{fig3:eps} represents synthetic images of our models (a),
(b), (c), and (d) at different frequency bands with the viewing angle
of ($\theta$, $\phi$)$=$($\pi/3$, $\pi/4$).  In Fig. \ref{fig3:eps},
upper half images correspond to the model with high-density plasma
($\dot{M}\approx 2.5\times 10^{-7} M_{\odot}\,{\rm yr}^{-1}$), whereas
lower half images correspond to that with low-density plasma
($\dot{M}\approx 2.5\times 10^{-8} M_{\odot}\,{\rm yr}^{-1}$).  Each
density model has two sub-categories; one is two-temperature plasma
$f_{\rm ep}=0.25$ and the other is one-temperature plasma $f_{\rm
  ep}=1$.  A basic feature of all images is that the core of emission
region is larger than the size of the event horizon shadow, and is
smaller than the radius of $50 r_{\rm s}$, except for the image of
model (b) (namely high-density and one-temperature plasma) at
millimeter band $10^{10} - 10^{11}$ Hz.

Distinctive feature of non-axisymmetry is seen in all models at
millimeter band.  The non-axisymmetric structure in the magnetized
accretion flows is also visible in Fig. \ref{fig1:eps}.  The apparent
size of emission region in models (a) and (d) looks similar, whereas
that in models (b) and (c) looks quite different.  This is because
the difference of density and temperature between model (a) and (d)
compensate each other.  On the other hand, non-axisymmetric feature
cannot be seen in sub-millimeter band ($10^{11} - 10^{12}$ Hz) except
for model (b), but it has nearly spherical emission region around the
gravitational centre. Again, the apparent size of emission region in
model (a) and (d) looks quite similar.  At both millimeter and
sub-millimeter bands, the core of emission region correspond to the
region of the strong magnetic field (see Fig. \ref{fig1:eps}c), and
the peak of emission region is slightly shifted from the gravitational
centre.

All models represent the disc-like structure in X-ray band $10^{17} -
10^{19}$ Hz.  The apparent size of emission region in high-density
models [(a) and (b)], and in low-density models [(c) and (d)] looks
similar each other.  This is because X-ray emission is produced by the
Bremsstrahlung emission, which is sensitive to density, not to
electron temperature.  Interestingly, asymmetric structure can be seen
only when electron temperature is equal to proton temperature [models
 (b) and (d)].  This feature is most likely to be produced by Compton
scattering of low-energy photons generated by synchrotron emission
(known as synchrotron self-Compton process: SSC).

\subsection{Spectral energy distribution (SED)}
\begin{figure*}
  \centerline{\epsfxsize=\hsize \epsfbox{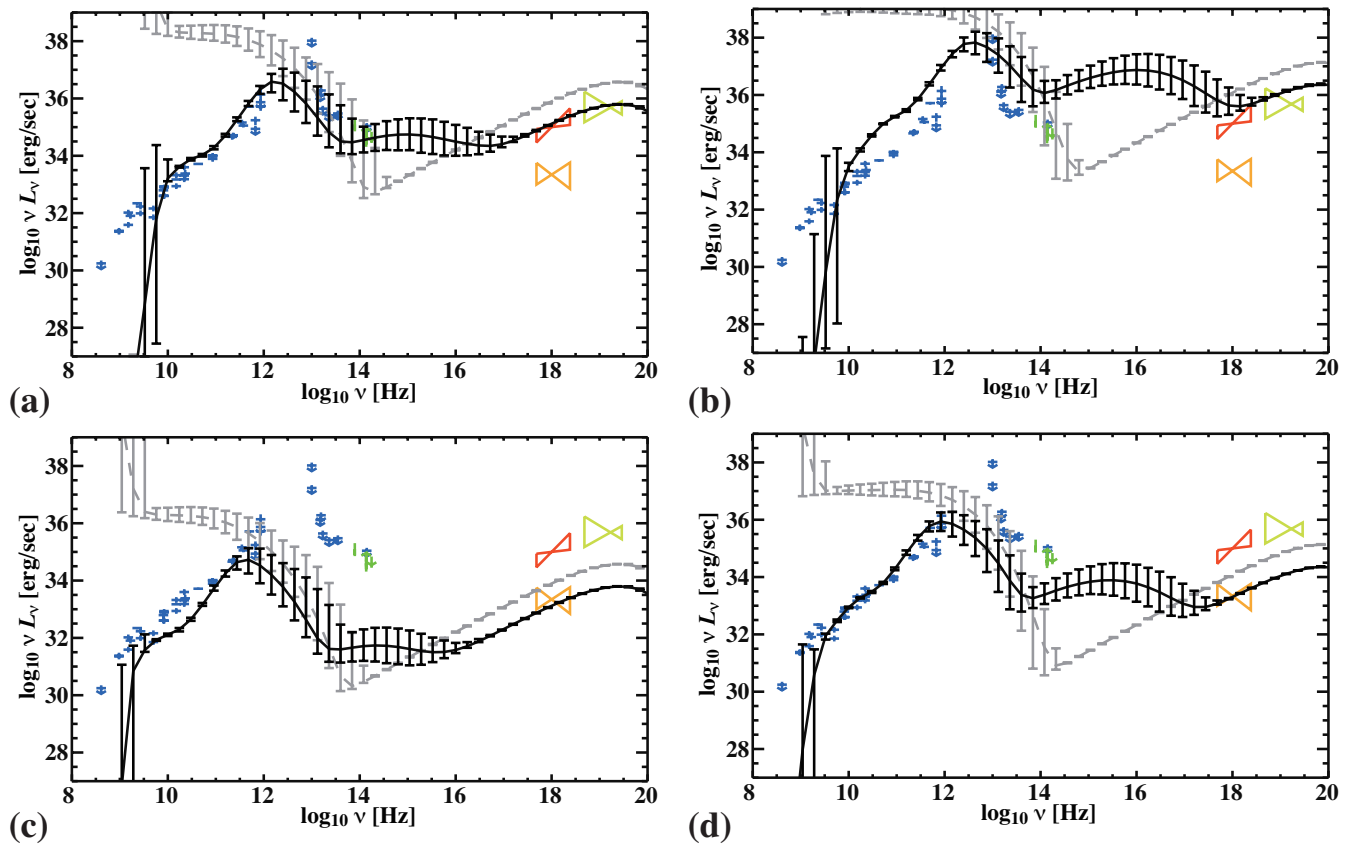}}
  \caption{Spectral energy distributions for a different plasma
  model.  (a) $\rho_{0}=8\times 10^{-15} {\rm [g\, cm^{-3}]}$ and
  $f_{\rm ep}=0.25$, (b) $\rho_{0}=8\times 10^{-15} {\rm [g\,
      cm^{-3}]}$ and $f_{\rm ep}=1$, (c) $\rho_{0}=8\times 10^{-16}
  {\rm [g\, cm^{-3}]}$ and $f_{\rm ep}=0.25$, and (d)
  $\rho_{0}=8\times 10^{-16} {\rm [g\, cm^{-3}]}$ and $f_{\rm ep}=1$.
  A black solid line and a grey dashed line indicate the emergent
  spectra and total emissivity spectra, respectively.  Black bars
  associated with solid and dashed lines indicate the time variation
  at a given frequency band.  Blue crosses and arrows indicate data
  taken by radio and infrared (IR) observations (Narayan et al. 1998
  and reference therein).  Green bars indicate data by near infrared
  observation (Eckart et al. 2006).  X-ray observations of the flaring
  and the quiescent state by {\it Chandra} are depicted as red and
  orange ``bow ties'', respectively (Baganoff et al. 2001) whereas
  gamma-ray observation of the flaring state by INTEGRAL is shown in
  light-green ``bow ties'' (Belandger et al. 2005).}
  \label{fig4:eps}
\end{figure*}

In addition to the spatial distributions of radiative fields,  we
calculate the SED of our MHD model.  In Fig. \ref{fig4:eps}, the
resultant SED is shown from radio to gamma-ray bands and also the
observed spectra of Sgr A* are superimposed.  The overall spectrum
consist of several radiative processes, that is, self-absorbed
synchrotron for $\nu\lsim 10^{12}$ Hz, synchrotron for
$10^{12}\lsim\nu\lsim 10^{14}$ Hz, synchrotron self-Compton for
$10^{14}\lsim\nu\lsim 10^{17}$ Hz, and thermal Bremsstrahlung for
$\nu\gsim 10^{17}$ Hz.  As we can see, time variability is different
at each band.  In optically thick region below the critical frequency
$\nu_{\rm c}\approx 10^{12}$ Hz, there is a small time variation. When
the frequency becomes higher, close to the critical frequency, the
time variation becomes larger.  Note that the variability at the
lowest frequency edge $\sim 10^{9}$ Hz depends on the number of photon
packets used in the calculation.  Because of the strong
self-absorption of synchrotron process, we cannot exclude the
statistical errors in those frequency bands.  Estimating from higher
frequency regions, the variability is about two factors in
magnitude. In optically thin region, on the other hand, time variation
becomes prominent in the low-frequency band $10^{12} - 10^{17}$ Hz
and it becomes negligible in high-energy band $\gsim 10^{17}$ Hz.

The resultant SED of model (a) can nicely fit the observed SED in
flaring state for all frequency range (Fig. \ref{fig4:eps}a).
The difference between the emergent spectra and the total emissivity
spectra below the critical frequency is induced by the synchrotron
self-absorption, whereas that in X-ray band is caused by the effect of
viewing angle.  Unlike the previous study of ADAF and OKM05,
low-frequency bump in the millimeter band $10^{10} - 10^{11}$ Hz is
reconstructed successfully without considering non-thermal electrons.
This is a main product of introducing a parameter $f_{\rm ep}$, which
can account for spatially structured electron temperature
distribution.

In order to explain two orders of magnitude difference in X-ray
emission between flaring and quiescent states, the density need to be
reduced by at least one order of magnitude, such as model (b) (see
Fig. \ref{fig4:eps}b).  The resultant SED of model (b) can fit the
X-ray spectra in quiescent state.  However it under-predicts the radio
to IR spectra for two-temperature plasma with $f_{\rm ep}=0.25$.
Remaining option is to increase electron temperature so as to
compensate the reduction of density, such as model (d) of
one-temperature plasma with $f_{\rm ep}=1$ (see
Fig. \ref{fig4:eps}d).  It turned out that the resultant SED of model
(d) can successfully reconstruct the X-ray spectra of quiescent state
as well as the radio and IR spectra.  Although the underlying physics
of electron heating is not clear, this is an outstanding result for
the theory of hot accretion flows.

It is interesting to note that the X-ray variability does not strongly
depend on the model parameter.  Actually, all models except model (b)
represent that the X-ray variability is very weak.  In model (b), soft
X-ray photons $\approx 10^{17}$ Hz are strongly contaminated by SSC
photons because both density and electron temperature are increased so
that scattering coefficient becomes large.  Moreover, it seems that
most of variabilities in ultraviolet band are produced by the
scattered photons and also they are correlated with IR variabilities
generated by synchrotron emission in all models. Note that the
amplitude of variabilities (different between the maximum and the
minimum power of escaping photons) in IR and UV bands does not change
much in the range of our model parameter.  Therefore, we suspect that
the variability in the recent NIR observation (Eckart et al. 2005) are
strongly affected by either the structure or the dynamics of the
accretion flows.

\section{Summary and Discussion}
We have investigated three-dimensional radiative features of
radiatively inefficient accretion flows modeled by MHD simulations.
The synthetic images of all models show that the core of emission
region in Sgr A* is larger than the size of the event horizon shadow,
and is smaller than the radius of $100\,r_{\rm s}$.  We have found
that a non-axisymmetric structure of $m=1$ is associated with an
elongated core emission in millimeter band.  Remarkably, the peak
location of the core emission in millimeter band $\approx 10^{11}$ Hz
is slightly shifted from the gravitational centre.  This is consistent
with the baseline-correlated flux density diagram of the recent VLBI
observation at 230 GHz (Doeleman et al. 2008).

We have also demonstrated for the first time that our 3-D MHD model
with different density (namely different mass accretion rate) can
reconstruct the observed broadband spectra including both the X-ray
quiescent and flaring states, simultaneously.  We have found that the
X-ray flaring state corresponds to relatively high-mass accretion rate
with a {\it weak} coupling between electron and proton temperature
$f_{\rm ep}=0.25$, whereas the X-ray quiescent state corresponds to
relatively low-mass accretion rate with a {\it strong} coupling
between electron and proton temperature $f_{\rm ep}=1$.  This is an
opposite sense if one considers only the Coulomb coupling for electron
heating.  We will discuss this issue in the following.

Heating mechanism of electron in hot accretion flows has been
investigated by many groups (Bisnovatyi-Kogan~\& Lovelace 1997;
Quataert 1998; Gruzinov 1998; Blackman 1999; Quataert~\& Gruzinov
1999; Medvedev 2000).  In ADAF/RIAF models, the turbulent viscosity
primarily heats protons and then hot protons heats electrons via
the Coulomb coupling between them.  When the proton temperature
becomes the virial temperature $\approx 10^{12}$ K, the
electron-proton Coulomb relaxation time becomes much larger than the
dynamical time (Spitzer 1956; Stepney 1983).  As a result, the
electron temperature decouple from the proton temperature.  This is a
well-known understanding of physics in hot accretion flows (Rees et
al. 1982; Narayan et al. 1995).  However, the spectra of hot accretion
flows modeled by MHD simulations indicate that a coupling ratio
$f_{\rm ep}$ is close to unity in both the flaring and quiescent state.
Moreover, $f_{\rm ep}$ increases when the mass accretion rate
decreases.  This is an opposite sense in terms of plasma physics,
because the electron-proton Coulomb relaxation time increases when the
density decrease as $t_{\rm relax}\propto T_{\rm
  e}^{3/2}/\rho_{0}$. Therefore, our results imply that the
alternative heating mechanism of electrons is requisite in order to
keep the electron temperature being close to the proton temperature.

Recently, SQHS07 have found that significant fraction of dissipative
energy generated by turbulent viscosity can be directly received by
electrons.  This can naturally explain the discrepancy of $f_{\rm ep}$
between our results and ADAF/RIAF models, because a fraction of energy
received by electron is assumed to be typically $\delta\sim m_{\rm
  e}/m_{\rm p}\sim 10^{-3}$ in ADAF models (e.g., Narayan et al .1998).
Although the detailed physics on the basis of viscous heating of
electrons is in dispute, their results are worth to implement as
subgrid physics of electron heating for 3-D global MHD simulations in
the near future.

\section*{Acknowledgments}
YK would like to thank S. Mineshige for numerous useful conversations
, and M. Miyoshi and M. Tsuboi for fruitful discussion on radio
spectrum of Sgr A*, and H. Hirashita and K. Yoshikawa for stimulating
discussions.  Radiation transfer code have been developed on {\it
  FIRST} simulator at the center for computational sciences,
University of Tsukuba.  Radiation transfer computations were carried
out on {\it FIRST} simulator and XT4 at the center for computational
astrophysics (CfCA), National Astronomical Observatory in Japan
(NAOJ).  MHD computations were carried out on VPP5000 at the
Astronomical Data Analysis Center of the National Astronomical
Observatory, Japan (yyk27b).  This work was supported in part by the
{\it FIRST} project based on Grants-in-Aid for Specially Promoted
Research by MEXT (16002003, MU), Grant-in-Aid for Scientific Research
(S) by JSPS (20224002, MU), and Grant-in-Aid for Scientific Research
of MEXT (20740115, KO).





\begin{thebibliography}{99}
\bibitem[\protect\citeauthoryear{Abramowicz}{1995}]{b4} Abramowicz,
  M.~A., Chen, X., Kato, S., Lasota, J.-P., \& Regev, O.\ 1995, ApJ,
  438, L37 

\bibitem[\protect\citeauthoryear{Baganoff}{2001}]{2001Natur.413...45B} Baganoff, F.~K., et al.\ 2001, Nature, 413, 45
\bibitem[\protect\citeauthoryear{Baganoff et al.}{2003}]{2003ApJ...591..891B} Baganoff, F.~K., et al.\ 2003, ApJ, 591, 891
\bibitem[\protect\citeauthoryear{Balbus}{1991}]{b15} Balbus, S.~A., \& Hawley, J.~F.\ 1991, ApJ, 376, 214
\bibitem[\protect\citeauthoryear{Balbus}{1998}]{b16} Balbus, S.~A., \& Hawley, J.~F.\ 1998, Reviews of Modern Physics, 70, 1
\bibitem[\protect\citeauthoryear{Balick}{1974}]{b7} Balick, B., \& Brown, R.~L.\ 1974, ApJ, 194, 265
\bibitem[\protect\citeauthoryear{Bisnovatyi-Kogan \& Lovelace}{1997}]{1997ApJ...486L..43B} Bisnovatyi-Kogan, G.~S., \& Lovelace, R.~V.~E.\ 1997, ApJ, 486, L43
\bibitem[\protect\citeauthoryear{Blackman}{1999}]{1999MNRAS.302..723B} Blackman, E.~G.\ 1999, MNRAS, 302, 723
\bibitem[\protect\citeauthoryear{Bower et al.}{2003}]{2003ApJ...588..331B} Bower, G.~C., Wright, M.~C.~H., Falcke, H., \& Backer, D.~C.\ 2003, ApJ, 588, 331
\bibitem[\protect\citeauthoryear{Bower}{2004}]{b9} Bower, G.~C., Falcke, H., Herrnstein, R.~M., Zhao, J.-H., Goss, W.~M., \& Backer, D.~C.\ 2004, Science, 304, 704
\bibitem[\protect\citeauthoryear{Bower et al.}{2005}]{2005ApJ...618L..29B} Bower, G.~C., Falcke, H., Wright, M.~C., \& Backer, D.~C.\ 2005, ApJ, 618, L29
\bibitem[\protect\citeauthoryear{Doeleman}{2008}]{b10} Doeleman,  S.~S., et al.\ 2008, Nature, 455, 78
\bibitem[\protect\citeauthoryear{Ghez}{2003}]{b12} Ghez, A.~M., et al.\ 2003, ApJ, 586, L127
\bibitem[\protect\citeauthoryear{Ghez}{2004}]{b13} Ghez, A.~M., et al.\ 2004, ApJ, 601, L159
\bibitem[\protect\citeauthoryear{Ghez}{2008}]{b13} Ghez, A.~M., et al.\ 2008, ApJ, 689, 1044
\bibitem[\protect\citeauthoryear{Gillessen}{2009}]{b14} Gillessen, S., Eisenhauer, F., Trippe, S., Alexander, T., Genzel, R., Martins, F., \& Ott, T.\ 2009, ApJ, 692, 1075
\bibitem[\protect\citeauthoryear{Gruzinov}{1998}]{1998ApJ...501..787G} Gruzinov, A.~V.\ 1998, ApJ, 501, 787
\bibitem[\protect\citeauthoryear{Hawley}{2000}]{b20} Hawley, J.~F.\ 2000, ApJ, 528, 462
\bibitem[\protect\citeauthoryear{Hawley}{2001}]{b18} Hawley, J.~F., \& Krolik, J.~H.\ 2001, ApJ, 548, 348
\bibitem[\protect\citeauthoryear{Ichimaru}{1977}]{b1} Ichimaru, S.\ 1977, ApJ, 214, 840
\bibitem[\protect\citeauthoryear{Igumenshchev}{2003}]{b22} Igumenshchev, I.~V., Narayan, R., \& Abramowicz, M.~A.\ 2003, ApJ, 592, 1042
\bibitem[\protect\citeauthoryear{Kato}{2004}]{b24} Kato, Y., Mineshige, S., \& Shibata, K.\ 2004, ApJ, 605, 307
\bibitem[\protect\citeauthoryear{Kato}{2004}]{b25} Kato, Y.\ 2004, PASJ, 56, 931
\bibitem[\protect\citeauthoryear{Kato}{2008}]{b6} Kato, S., Fukue, J., \& Mineshige, S.\ 2008, Black-Hole Accretion Disks --- Towards a New Paradigm ---, 549 pages, including 12 Chapters, 9 Appendices,  ISBN 978-4-87698-740-5, Kyoto University Press (Kyoto, Japan), 2008.,
\bibitem[\protect\citeauthoryear{Loeb \& Waxman}{2007}]{2007JCAP...03..011L} Loeb, A., \& Waxman, E.\ 2007, Journal of Cosmology and Astro-Particle Physics, 3, 11
\bibitem[\protect\citeauthoryear{Matsumoto}{1999}]{b19} Matsumoto, R.\ 1999, Numerical Astrophysics, 240, 195
\bibitem[\protect\citeauthoryear{Matsumoto}{1998}]{1998} Matsumoto, M. \& Nishimura, T.\ 1998, ACM Trans. on Modeling and Computer Simulation, 8, 3
\bibitem[\protect\citeauthoryear{Machida}{2000}]{b21} Machida, M., Hayashi, M.~R., \& Matsumoto, R.\ 2000, ApJ, 532, L67
\bibitem[\protect\citeauthoryear{Mahadevan}{1998}]{1998Natur.394..651M} Mahadevan, R.\ 1998, Nature, 394, 651
\bibitem[\protect\citeauthoryear{Moscibrodzka}{2007}]{b5} Moscibrodzka, M., Proga, D., Czerny, B., \& Siemiginowska, A.\ 2007, A\&A, 474, 1 (MPCS07)
\bibitem[\protect\citeauthoryear{Medvedev}{2000}]{2000ApJ...541..811M} Medvedev, M.~V.\ 2000, ApJ, 541, 811
\bibitem[\protect\citeauthoryear{Narayan}{1994}]{b2} Narayan, R., \&
  Yi, I.\ 1994, ApJ, 428, L13
\bibitem[\protect\citeauthoryear{Narayan}{1995}]{b3} Narayan, R., Yi, I., 
\& Mahadevan, R.\ 1995, Nature, 374, 623

\bibitem[\protect\citeauthoryear{Ohsuga}{2005}]{b1} Ohsuga, K., Kato, Y., \& Mineshige, S.\ 2005, ApJ, 627, 782 (OKM05)
\bibitem[\protect\citeauthoryear{Pacholczyk}{1970}]{1970ranp.book.....P} Pacholczyk, A.~G.\ 1970, Series of Books in Astronomy and Astrophysics, San Francisco: Freeman, 1970,
\bibitem[\protect\citeauthoryear{Proga}{2003}]{b23} Proga, D., \& Begelman, M.~C.\ 2003, ApJ, 592, 767
\bibitem[\protect\citeauthoryear{Pozdnyakov et al.}{1977}]{1977Astron} Pozdnyakov, L.~A., Sobol, I.~M., \& Syunyaev, R.~A.\ 1977, Sov. Astron., 21, 708
\bibitem[\protect\citeauthoryear{Quataert}{1998}]{1998ApJ...500..978Q}
  Quataert, E.\ 1998, ApJ, 500, 978
\bibitem[\protect\citeauthoryear{Quataert \& Gruzinov}{1999}]{1999ApJ...520..248Q} Quataert, E., \& Gruzinov, A.\ 1999, ApJ, 520, 248
\bibitem[\protect\citeauthoryear{Rees et al.}{1982}]{1982Natur.295...17R} Rees, M.~J., Begelman, M.~C., Blandford, R.~D., \& Phinney, E.~S.\ 1982, Nature, 295, 17
\bibitem[\protect\citeauthoryear{Rybicki \& Lightman}{1979}]{1979rpa..book.....R} Rybicki, G.~B., \& Lightman, A.~P.\ 1979, New York, Wiley-Interscience, 1979.~393 p.,
\bibitem[\protect\citeauthoryear{Sch{\"o}del et al.}{2002}]{2002Natur.419..694S} Sch{\"o}del, R., et al.\ 2002, Nature, 419, 694
\bibitem[\protect\citeauthoryear{Sharma et al.}{2007}]{2007ApJ...667..714S} Sharma, P., Quataert, E., Hammett, G.~W., \& Stone, J.~M.\ 2007, ApJ, 667, 714
\bibitem[\protect\citeauthoryear{Shen}{2005}]{b8} Shen, Z.-Q., Lo, K.~Y., 
Liang, M.-C., Ho, P.~T.~P., \& Zhao, J.-H.\ 2005, Nature, 438, 62
\bibitem[\protect\citeauthoryear{Spitzer}{1956}]{1956pfig.book.....S} Spitzer, L.\ 1956, Physics of Fully Ionized Gases, New York: Interscience Publishers, 1956,
\bibitem[\protect\citeauthoryear{Stepney}{1983}]{1983MNRAS.202..467S} Stepney, S.\ 1983, MNRAS, 202, 467
\bibitem[\protect\citeauthoryear{Stepney \& Guilbert}{1983}]{1983MNRAS.204.1269S} Stepney, S., \& Guilbert, P.~W.\ 1983, MNRAS, 204, 1269
\bibitem[\protect\citeauthoryear{Stone}{2001}]{b17} Stone, J.~M., \& Pringle, J.~E.\ 2001, MNRAS, 322, 461
\bibitem[\protect\citeauthoryear{Yuan}{2003}]{b5} Yuan, F., Quataert,
  E., \& Narayan, R.\ 2003, ApJ, 598, 301
\end{thebibliography}
\end{document}